\documentclass[11pt]{article}
\usepackage[toc,page]{appendix}
\RequirePackage{fix-cm}
\usepackage[utf8]{inputenc}
\usepackage[margin=1.5in]{geometry}
\usepackage[bottom, flushmargin]{footmisc}
\usepackage{marvosym}
\usepackage{titlesec}

\titlespacing\section{0pt}{0.21in}{0.03in}
\titlespacing\subsection{0pt}{0.21in}{0.01in}
\titlespacing\subsubsection{0pt}{0.21in}{0in}
\usepackage{marginnote}

\usepackage[superscript]{cite}
\usepackage{enumitem} 
\newcommand{\bulletlabel}{\raisebox{0.2ex}{\small$\bullet$}}
\newlist{enumbul}{enumerate}{6}
\setlist[enumbul]{label=\bulletlabel, topsep=0pt, leftmargin=0.3in, rightmargin=0.3in}
\newlist{enumar}{enumerate}{6}
\setlist[enumar]{label=\arabic*$)$,topsep=0pt, leftmargin=0.3in, rightmargin=0.3in}

\usepackage{hyperref}
\usepackage{comment} 
\usepackage{enumitem} 
\usepackage{graphicx,nicefrac} 
\usepackage[justification=centering]{caption}
\graphicspath{ {dataviz/} }
\usepackage{amsmath} 
\usepackage{amssymb} 
\usepackage{amsthm}
\usepackage{bm}
\usepackage{float}
\usepackage{mathtools}
\usepackage{bigints}

\usepackage[most]{tcolorbox}
\usepackage{color} 
\usepackage{xcolor, colortbl} 

\hypersetup{
	colorlinks,
	linkcolor=black, 
	urlcolor=cyan,  
	citecolor= red, 
	breaklinks=true
}

\usepackage{multirow}
\usepackage{multicol}
\usepackage{array}
\usepackage{booktabs}
\newcolumntype{L}[1]{>{\raggedright\let\newline\\\arraybackslash\hspace{0pt}}m{#1}}
\newcolumntype{C}[1]{>{\centering\let\newline\\\arraybackslash\hspace{0pt}}m{#1}}
\newcolumntype{R}[1]{>{\raggedleft\let\newline\\\arraybackslash\hspace{0pt}}m{#1}}

\begin{document}
\pagenumbering{gobble}
\def\acronymend{MECo}
\def\acronym{\acronymend\space}

\def\TOTALBALLOTS{25,545}
\def\TOTALCONTESTS{9,704}
\def\TOTALPARTIES{110}
\def\TOTALCOALITIONS{18}
\def\TOTALCONTESTSSTATE{6,936}
\def\TOTALCONTESTSFEDERAL{2,715}
\def\TOTALCONTESTSBYELEC{53}
\def\TOTALELECTIONS{16}
\def\UNIQUECANDIDATES{14,000}
\def\TOTALCORRECTIONS{114}

\phantom{Thevesh}

\vfill

\begin{center}
	\begin{minipage}{0.88\linewidth}
		\begin{center}
			\LARGE
			The Malaysian Election Corpus (\acronymend):\\Electoral Maps and Cartograms from 1954 to 2025\\[0.21in]
			\Large
			Thevesh Thevananthan\footnotemark$^{\text{\Letter}}$ and Danesh Prakash Chacko\footnotemark\\[0.15in]
		\end{center}
		\large
		Electoral boundaries in Malaysia are not publicly available in machine-readable form. This prevents rigorous analysis of geography-centric issues such as malapportionment and gerrymandering, and constrains spatial perspectives on electoral outcomes. We present the second component of the Malaysian Election Corpus (MECo), an open-access collection of digital electoral boundaries covering all 19 approved delimitation exercises in Malaysia's history, from the first set of Malayan boundaries in 1954 until the 2019 Sabah delimitation. We also auto-generate election-time maps for all federal and state elections up to 2025, and include equal-area and electorate-weighted cartograms to support deeper geospatial analysis. This is the first complete, publicly-available, and machine-readable record of Malaysia's electoral boundaries, and fills a critical gap in the country's electoral data infrastructure.
	\end{minipage}
\end{center}

\footnotetext[1]{Malaya University, W.P. Kuala Lumpur, Malaysia\\\Letter\space \href{mailto:thevesh.theva@gmail.com}{thevesh.theva@gmail.com}}
\footnotetext[2]{TindakMalaysia Network Services PLT (``Tindak Malaysia'')}

\vfill

\phantom{Word Count:}

\newpage
\pagenumbering{arabic}

\section*{Background \& Summary}
Electoral boundaries provide the operational framework through which democratic representation is organised, particularly in countries that employ first-past-the-post (FPTP) constituency-based electoral systems. Boundaries determine how populations are grouped, how votes are aggregated, and how representation is translated into seats. As a result, almost all spatial analyses of elections---covering topics such as malapportionment, gerrymandering, voter access, and constituency change---depend fundamentally on the availability and quality of boundary data. When such data are missing, incomplete, or inconsistent across time, electoral outcomes cannot be reliably mapped, and changes in constituency composition become difficult to trace and study systematically.

In democracies with well-established norms of transparency, such as the United Kingdom, Australia, and New Zealand, electoral boundaries are treated as core public data infrastructure. Election commissions, census bureaus, or national statistical offices routinely publish constituency boundaries in machine-readable formats, enabling direct linkage to election results and other datasets. The availability of such data has underpinned a large body of empirical work in political science and political geography, including research on malapportionment and spatial electoral inequality.\cite{openshaw1984modifiable,samuels2001value} More broadly, electoral maps are no longer treated as ancillary visual material, but rather as first-class components of election data infrastructure designed to ensure seamless interoperability with tabular election results.\cite{clea}

\begin{figure*}[htb]
	\centering
	\caption{Parliaments by state after each round of delimitation}
	\vspace{0in}
	\makebox[\textwidth][c]{
		\includegraphics[width=1.1\linewidth]{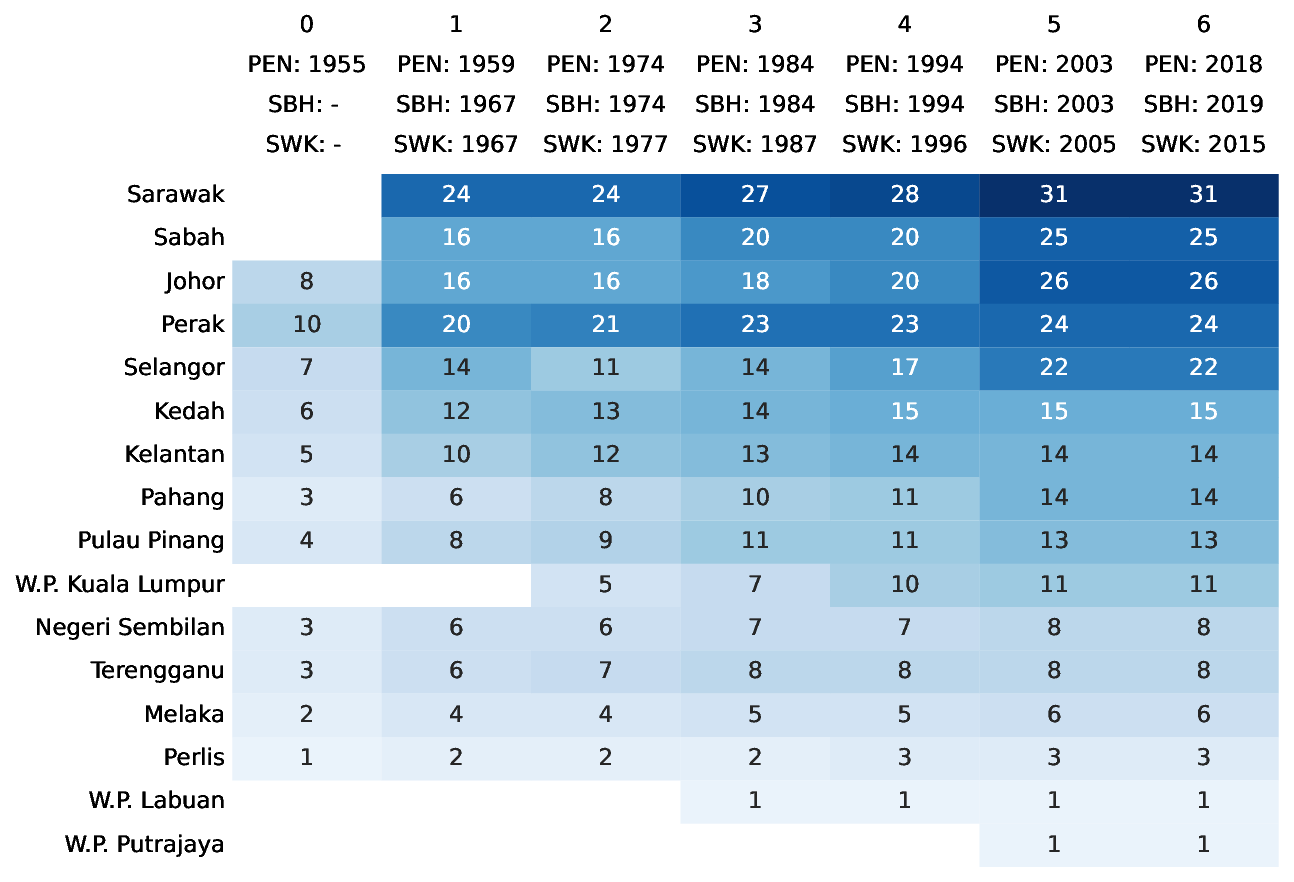}
	}
	\label{fig:nparlimen}
\end{figure*}

\begin{figure*}[htb]
	\centering
	\caption{DUN constituencies by state after each round of delimitation}
	\vspace{0in}
	\makebox[\textwidth][c]{
		\includegraphics[width=1.1\linewidth]{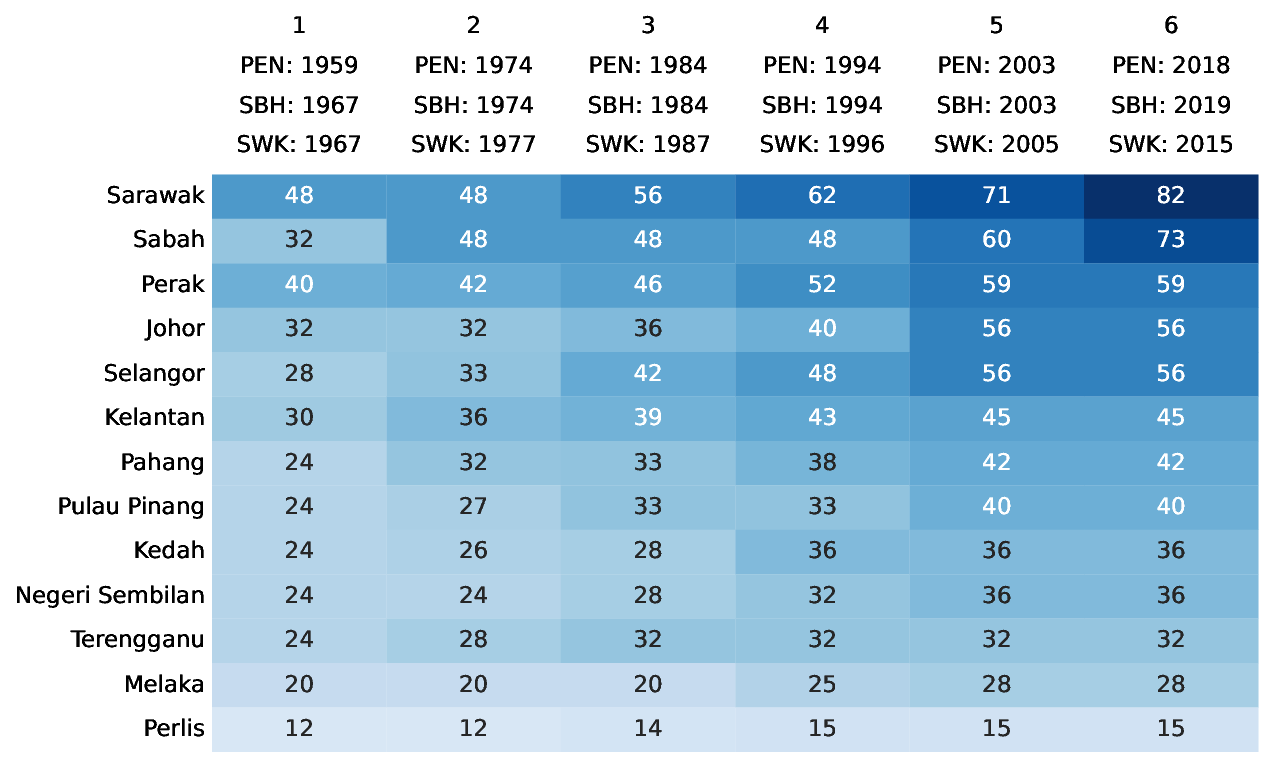}
	}
	\label{fig:ndun}
\end{figure*}

Malaysia presents a contrasting case. Under the Federal Constitution, the Election Commission of Malaysia (EC) conducts delimitation exercises separately for three geographic units of review---Peninsular Malaysia (States of Malaya; including all Federal Territories), Sabah, and Sarawak---and submits its recommendations to the Prime Minister, who may then table the report for approval by the Dewan Rakyat (House of Representatives). To date, there have been 7 approved delimitations for Peninsular Malaysia (1 for Malaya before independence), 6 for Sabah, and 6 for Sarawak (Figures \ref{fig:nparlimen} and \ref{fig:ndun}). However, while approved delimitations must be gazetted, the governing framework does not require the gazette to include a cartographic representation of constituency boundaries (the Gazetted Plan) as part of the permanent public record.

Compounding this limitation, the EC has never openly published constituency boundaries in a structured, reusable format. Earlier practice---up to the federal general election in 1990 and the Sabah state election in 1999---did include printed boundary maps in official post-election reports, but this convention was later discontinued. As a result, the public record of Malaysia's electoral geography has become increasingly fragmented over time. Furthermore, it is doubtful as to whether even the EC possesses historical boundaries in digital form; the geospatial data they sell covers the most recent set of delimitations only. The cumulative effect is that Malaysia's electoral geography has remained difficult to access for systematic research, despite the central role that boundary change plays in debates over representation and electoral fairness.

There have been important local efforts to address parts of this gap. Civil society and media organisations---notably Tindak Malaysia---have produced maps and visual materials to support public scrutiny of redelineation exercises. These initiatives have played an important role in political engagement and advocacy. However, they were not designed to function as a comprehensive research archive: they typically focus on recent delimitations, employ heterogeneous formats and conventions, and are not subjected to systematic validation. As a result, they cannot readily support longitudinal or comparative analysis of boundary change since independence.

In this paper, we address this gap by providing the second component of the Malaysian Election Corpus (MECo),\cite{thevesh2025meco} namely the first complete, publicly-available, and machine-readable collection of Malaysia's electoral boundaries, covering all 19 approved delimitations from the first set of Malayan boundaries in 1954 up to the most recent delimitation of Sabah in 2019. Except for the 1954 delimitation which was in effect only for the 1955 Federal Legislative Council election, all boundary sets include both parliamentary (Parliament) and state (DUN) constituency boundaries. In addition to raw boundaries, we also include equal-weight and electorate-weighted Dorling cartograms to support comparative analysis and visualisation, thus following a well-established cartographic literature that recognises the limits of conventional maps for representing population-weighted phenomena.\cite{dent1975communication}

Crucially, our collection was designed from the outset as a complement to the first component of MECo,\cite{thevesh2025meco} which provides a complete and standardised record of Malaysian federal and state election results. As a living resource, this collection lowers or even eliminates the fixed cost of working with Malaysia's electoral geography; we hope this will enable a new wave of research on malapportionment, gerrymandering, constituency lineage, and the spatial dynamics of political competition.

Finally, we note that while this collection is novel for Malaysia, it follows a growing body of recent work\cite{taylor2023canada, lewis2013us} focused on compiling and curating country-specific electoral geography for reuse as open research infrastructure. However, even in countries with robust public data ecosystems, scholars often face substantial challenges in fully reconstructing historical boundary sets, as earlier delimitations frequently exist in mediums which were never intended for long-term preservation. By documenting this reconstruction process, we illustrate a practical pathway for digitising and validating legacy electoral boundaries, an approach that may be applied to other national contexts where historical electoral geography remains inaccessible.

\section*{Methods}

We digitised electoral boundaries for all 19 approved delimitations in Malaysia's history, covering all federal parliamentary constituencies (`Parliaments') and state legislative constituencies (`DUNs'). We digitised these boundaries using QGIS, an industry-grade Geographic Information System (GIS) software which supports rasterization and geo-referencing.

For the 13 delimitation exercises before 2000, we traced the boundaries based on official constituency maps published by the EC in physical election reports or delimitation reports which we scanned; for the 6 delimitation exercises after 2000, we relied on PDFs or images of maps produced by the EC. Although the EC does not systematically disseminate or make this source material available to the general public, it is legally classified as open data (Terbuka) and is not copyrightable under Malaysian law. Section 3(1) of the Copyright Act 1987 (Act 332) expressly excludes from copyright ``official texts of the Government or statutory bodies of a legislative or regulatory nature''. The gazetted election boundaries fall under this category. This interpretation was confirmed through consultation with legal advisors familiar with Malaysian election documentation and copyright law.

The most important point for users to note is that our goal---as is common in the GIS literature dealing with historical boundaries for analytical use\cite{knowles2002past, long1995atlas}---was to reconstruct constituency boundaries to support research and public communication, rather than to reproduce legally operative boundaries at cadastral precision. In simple language, these boundaries are suitable for  spatial analysis, visualisation, and longitudinal comparison, but are not suitable or intended for operational or legal uses such as assigning individual voters to constituencies.

\subsection*{Tracing and labelling}

In Malaysia, all DUNs geographically reside strictly within one Parliament, even though there is no reporting line between state and federal-level representatives. Therefore, our strategy was to digitise the boundaries at Parliament level, and subsequently subdivide each Parliament into its constituent DUNs. The only exception to this strategy was for the 1954 delimitation of Malayan boundaries, where constituencies were not sub-divided because they were intended to be used only for the pre-independence Federal Legislative Council election of 1955; for this case only, federal boundaries were traced directly.

To digitise the boundaries, we first rasterised our source images and PDFs, and then geo-referenced them against contemporary basemaps. There are three important facts justifying our use of contemporary basemaps:
\begin{enumar}
	\item Legally, electoral boundaries cannot cross state boundaries. In short, every Parliament and DUN resides strictly within one state.
	\item With the exception of the carving out of the three Federal Territories---Kuala Lumpur in 1974, Labuan in 1984, and Putrajaya in 2001---Malaysia's state boundaries have not been altered since independence.
	\item The numerous documented instances of land reclamation and coastal changes since 1954 only affect the `edges' of states, but not the divisions between states.
\end{enumar}

Put together, this allowed us the convenience of using contemporary basemaps to establish approximate spatial alignment; we would otherwise have had to accept the historical maps `as-is', with state boundaries derived as the aggregation of DUN and Parliament boundaries.
Once we rasterised and geo-referenced our source material, we traced the boundaries by hand, following the midline of the printed boundary representation. Our goal during this process was to ensure that overall shapes, relative positions, and adjacency relationships were preserved to the highest degree of accuracy, thus guaranteeing topological and analytical correctness. However, we intentionally avoided unnecessary geometric complexity, such as tracing every minor curve in a river, or every jagged edge of a mountain range. This approach is consistent with common practices in analysis-oriented boundary datasets, especially for mountainous tropical countries like Malaysia, where the size of geospatial data can increase by a full order of magnitude if total accuracy of natural features is a strict requirement.

Once digitisation was complete, each polygon was assigned a consistent set of hierarchical identifiers. Parliaments were linked to their corresponding state, while DUNs were linked to both their parent Parliament and state. This explicit encoding of parent-child relationships allows boundaries to be aggregated, queried, and validated programmatically, and ensures direct interoperability with the constituency identifiers used in election results datasets.

\subsection*{Accounting for land reclamation and erosion}

Malaysia has experienced substantial land reclamation and coastal changes since independence, particularly in urban and industrialised coastal areas such as Penang, Johor Bahru, and the Klang Valley. Because electoral boundaries are defined with reference to geography at the time of delimitation, we digitised boundaries with respect to the historical coastline and land extent implied by the original maps, rather than forcing alignment with contemporary shorelines.

\begin{figure*}[htb]
	\centering
	\caption{Implied coastline (aggregation of DUNs) of Penang Island}
	\vspace{0in}
	\makebox[\textwidth][c]{
		\includegraphics[width=0.6\linewidth]{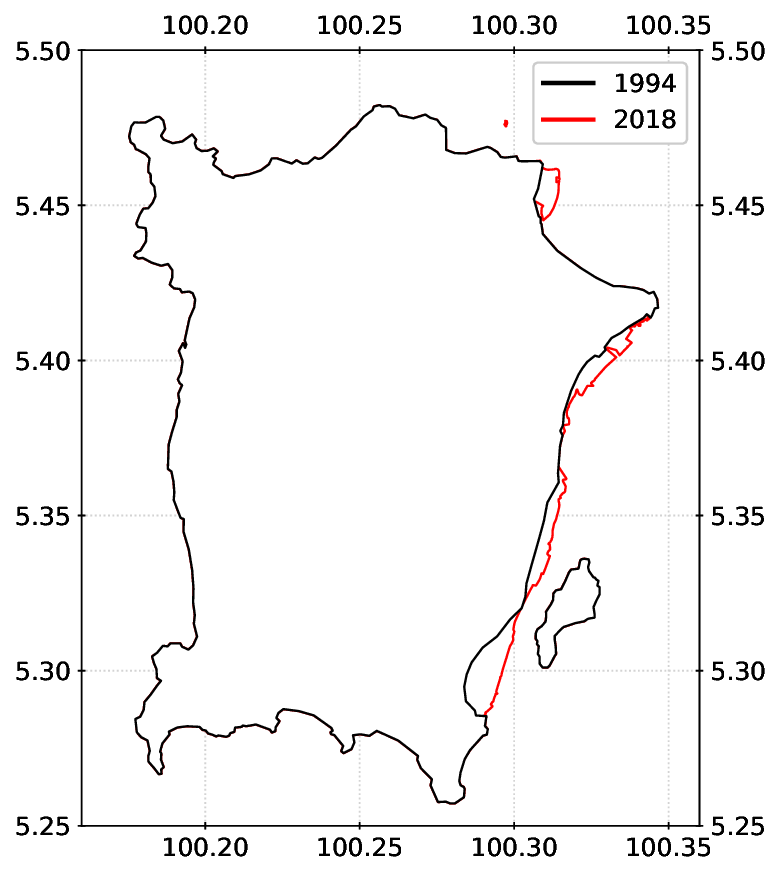}
	}
	\label{fig:penang}
\end{figure*}

The case of Penang illustrates this approach. Large areas of reclaimed land along the island's eastern coast post-date several delimitation exercises (Figure \ref{fig:penang}). In these cases, we traced constituencies according to the boundary lines depicted in the original EC maps, even where those lines now fall offshore or intersect reclaimed areas. This ensures that historical boundaries remain faithful to their original intent and avoids retroactively distorting earlier delimitations using present-day geography. Consequently, although we do not intend for users to employ our maps for research on coastal change, discrepancies between historical boundaries and current coastlines should generally be interpreted as artefacts of physical change, rather than digitisation error.

\subsection*{Accounting for islands}

\begin{figure*}[htb]
	\centering
	\caption{P.189 Semporna as per the 2019 Sabah delimitation}
	\vspace{0in}
	\makebox[\textwidth][c]{
		\includegraphics[width=0.67\linewidth]{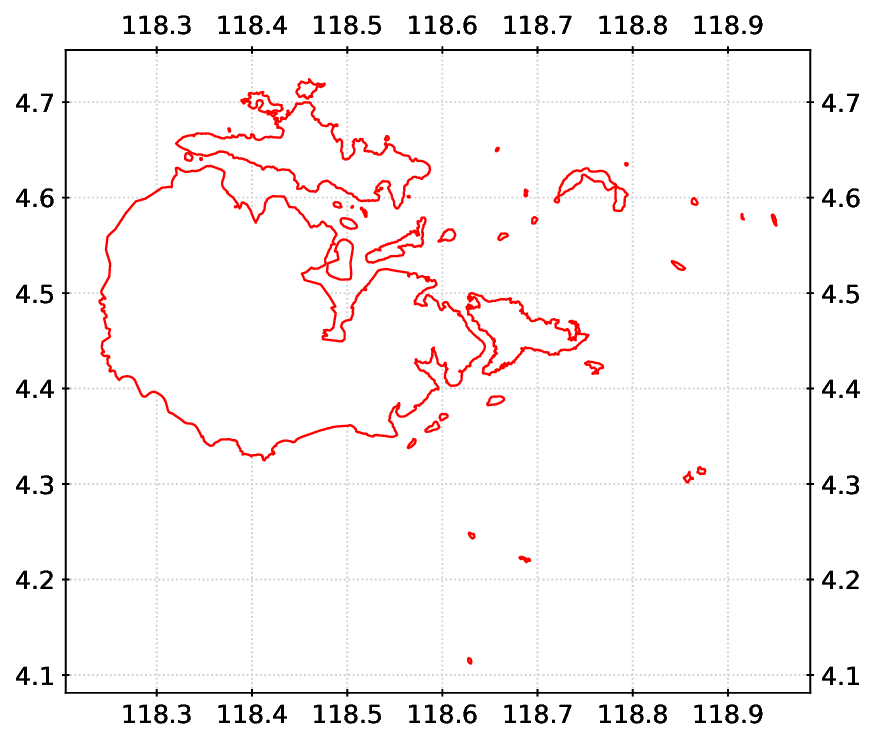}
	}
	\label{fig:semporna}
\end{figure*}

For contiguous landmasses, each sub-division is a single, closed geometry. This is not the case for Malaysia, due to the presence of 879 islands surrounding its two main landmasses (Peninsular and East Malaysia), with the largest number for a single state located in the waters of Sabah. Specifically, 48 out of 222 Parliaments and 70 out of 600 DUNs are non-contiguous due to islands; at the extreme, the Parliament of Semporna comprises its main landmass and approximately 56 surrounding islands (Figure \ref{fig:semporna}).

This is not an issue from a technical perspective, as modern GIS software can handle the presence of what is called MultiPolygons. However, correctly assigning identifiers to island components required careful manual verification, especially for historical maps which often omit small islands. To handle this, we cross-referenced our source material with gazette descriptions where available, and inferred the parent landmass via spatial proximity if absolutely necessary. All island geometries were assigned the same constituency identifiers as their associated mainland polygon, ensuring that each Parliament or DUN is represented as a single logical unit regardless of physical contiguity. That having been said, we make room for the possibility that some minor islands---especially unpopulated ones not represented on any electoral maps---are omitted in totality from our collection. As such, we do not intend for users to use our maps for studying Malaysian islands from a purely geographic perspective.

\subsection*{Election-specific boundary generation}

Delimitation exercises in Malaysia are conducted independently for Sabah, Sarawak, and Peninsular Malaysia, and do not occur synchronously across these three units of review. As a result, a single general election is conducted using a combination of delimitation vintages in effect at the time of polling. For example, the 2018 federal general election was conducted using the 2018 Peninsular Malaysia delimitation, the 2015 Sarawak delimitation, and the 2003 Sabah delimitation.

To avoid users having to manually account for this, we programmatically generated an election-specific boundary set for each election, which selects the exact constituency boundaries that were legally in force in each region at the time of polling. This design ensures that election results can be analysed directly against the correct operative boundaries, and avoids common sources of downstream error when working with Malaysian election geography. It also preserves a clear separation between delimitation exercises and elections, allowing users to study either independently while retaining an explicit and automated mapping between the two.

\subsection*{Cartogram generation}

\begin{figure*}[htb]
	\centering
	\captionsetup{justification=centering}
	\caption{GE-15 winning coalitions in Peninsular Malaysia, visualised using a choropleth map vs a Dorling cartogram}
	\vspace{0in}
	\makebox[\textwidth][c]{
		\includegraphics[width=1.2\linewidth]{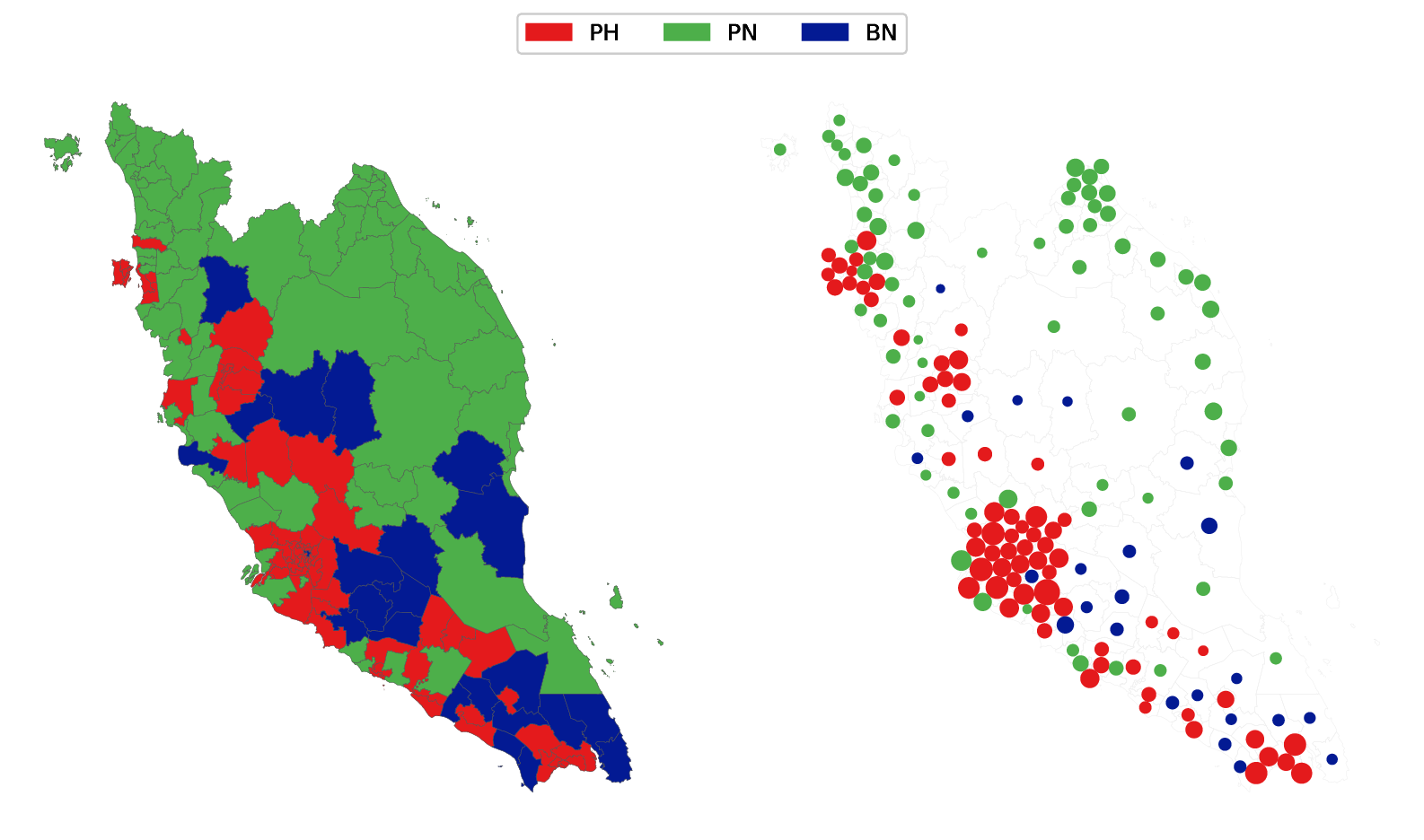}
	}
	\label{fig:choropleth_v_dorling}
\end{figure*}

In addition to geographic constituency boundaries, we generated two classes of derived cartographic representations for each election map: equal-weight Dorling cartograms and electorate-weighted Dorling cartograms.\cite{dorling2011area} In the equal-weight cartograms, all constituencies have the same area, thus enabling visual representations that reflect formal legislative equality (i.e.\ each Member has one vote in Parliamentary or DUN proceedings). In the electorate-weighted cartograms, constituency area is scaled in proportion to the size of the electorate, thus accurately representing the true incidence of electoral outcomes (Figure \ref{fig:choropleth_v_dorling}). For both types, a key step in the process was to decide on the expansion factor---typically denoted as the \textit{k} parameter---used to parameterise the tradeoff between maximising the visual field and preserving spatial correspondence (specifically, the position of each constituency's centroid). We found it more effective to choose \textit{k} via trial and error rather than programmatic methods, especially since there is no single correct method to derive an optimal \textit{k} value as a closed-form solution. In short, we iteratively inspected our generated cartograms, and fine-tuned the value of \textit{k} used for each one to ensure a consistent visual feel, especially as the number of constituencies increased over time.

\begin{figure*}[htb]
	\centering
	\captionsetup{justification=centering}
	\caption{GE-15 winning coalitions in Peninsular Malaysia, visualised using a topology-preserving cartogram}
	\vspace{0in}
	\makebox[\textwidth][c]{
		\includegraphics[width=0.6\linewidth]{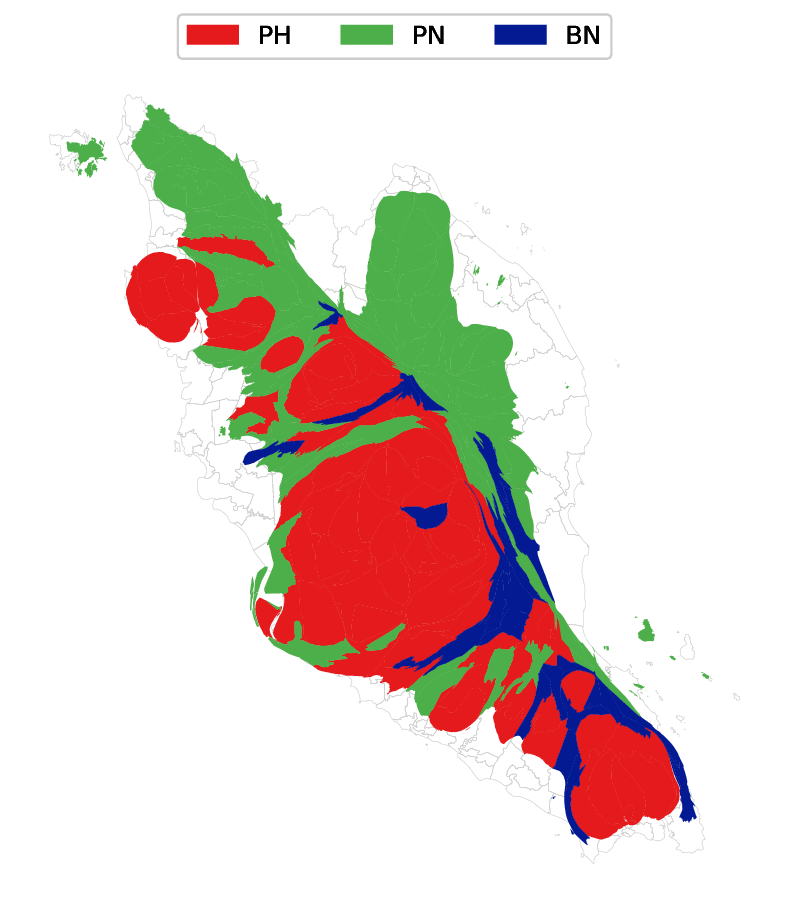}
	}
	\label{fig:topology_preserving_cartogram}
\end{figure*}

While there are many methods for generating cartograms, each with their own advantages and disadvantages,\cite{nusrat2016state} we argue that the Dorling cartogram is well-suited to the Malaysian electoral context. This is primarily due to extreme outliers at both ends of the density-area spectrum: the central Klang Valley comprises roughly 1\% of Malaysia's land area but over 25\% of its voters, while Sarawak accounts for nearly 40\% of Malaysia's land area but just 6\% of its voters. Under such conditions, shape-preserving (isomorphic) cartograms necessarily allocate roughly 96\% of the map to empty space, while topology-preserving cartograms can heavily distort the shape and position of constituencies (Figure \ref{fig:topology_preserving_cartogram}). By contrast, Dorling cartograms---which are also well-suited to plotting in conjunction with the original map as a base---intuitively illustrate normalised electoral outcomes while preserving approximate spatial location.

That having been said, our approach does not constrain users who wish to use a different cartogram-generation algorithm; to do so, users need only implement their algorithm of choice on the delimitation or election-time maps. Furthermore, users should note that these cartograms are provided for analytical and visualisation purposes only; because they do not preserve adjacency relationships, they should not be interpreted as geographic boundaries. For electorate-weighted cartograms in particular, we emphasise that they are best used to communicate electoral outcomes, rather than to illustrate malapportionment, which is more transparently conveyed using tabular or bar-chart representations of constituency size.

\section*{Data Records}

All datasets (Table~\ref{tab:data_records}) are published on Harvard Dataverse,\cite{meco2_dataset} which serves as the canonical archive. For convenience, the exact same datasets are also mirrored on:

\begin{enumbul}
	\item \textbf{Zenodo}\cite{meco2_codebase} (\href{https://doi.org/10.5281/zenodo.18093017}{doi.org/10.5281/zenodo.18093017})\\
	Provides code and raw source files, in addition to datasets.

	\item \textbf{GitHub} (\href{https://github.com/Thevesh/paper-meco-maps}{github.com/Thevesh/paper-meco-maps})\\
	Facilitates active development and maintenance, issue tracking and community contributions. Substantial updates are released on Zenodo.
\end{enumbul}

\begin{table}[!htb]
	\centering
	\renewcommand{\arraystretch}{1.05}
	\caption{Description of primary datasets}
	\label{tab:data_records}
	\vspace{0.04in}
	\begin{tabular}{L{0.5\linewidth}L{0.5\linewidth}}
		\toprule
		\textbf{Filename}                                                                   & \textbf{Description}                                                                                                                            \\
		\midrule
		\texttt{\{format\}/delimitations/*}\linebreak                                       & Maps corresponding to every delimitation exercise in Malaysia's history, 19 in total                                                            \\[0.17in]
		\texttt{\{format\}/elections/*}\linebreak\phantom{a}\linebreak\phantom{a}\linebreak & Maps corresponding to every federal (16 in total) and state election (192 in total) in Malaysia's history, compiled from the base delimitations \\[0.43in]
		\texttt{\{format\}/cartogram-electorate/*}\linebreak\phantom{a}\linebreak           & Electorate-weighted cartograms, with one generated for each election-specific map within \texttt{/elections}                                    \\[0.3in]
		\texttt{\{format\}/cartogram-equal/*}\linebreak\phantom{a}\linebreak                & Equal-weight cartograms, with one generated for each election-specific map within \texttt{/elections}                                           \\[0.3in] \hline
		                                                                                    &                                                                                                                                                 \\[-0.14in]
		\texttt{delims\_to\_elections.csv}\linebreak                                        & Specification of the delimitations used to generate each election map                                                                           \\[0.17in]
		\texttt{cartogram\_electorate\_k.csv}\linebreak\phantom{a}\linebreak                & Specification of the manually-tuned \textit{k} parameter used to generate the electorate-weighted cartograms                                    \\[0.3in]
		\texttt{cartogram\_equal\_k.csv}\linebreak\phantom{a}\linebreak                     & Specification of the manually-tuned \textit{k} parameter used to generate the equal-weight cartograms                                           \\[0.3in]
		\bottomrule
	\end{tabular}
\end{table}

To maximise usability across research, engineering, GIS, and public-facing applications, all boundary datasets are distributed in five complementary open formats, each optimised for a distinct class of workflows:
\begin{enumbul}
	\item \textbf{GeoJSON} \texttt{(.geojson)}\\
	GeoJSON is the most widely supported and human-readable vector format for geospatial data. It is natively supported by nearly all GIS software, web-mapping libraries, and scripting environments. GeoJSON serves as the lowest-barrier entry point for users and is particularly well suited for lightweight analysis, inspection, and browser-based visualisation.
	\item \textbf{TopoJSON} \texttt{(.topojson)}\\
	TopoJSON is a topology-aware extension of GeoJSON that encodes shared boundaries only once. This results in substantial file size reduction and, critically, guarantees perfect alignment between adjacent constituencies, eliminating even the possibility of slivers or gaps that can arise from independent polygon representations. TopoJSON is particularly valuable for cartography or any workflow where topological consistency is important.
	\item \textbf{GeoParquet} \texttt{(.geoparquet)}
	GeoParquet is the primary format intended for analytical and data-engineering workflows. Built on Apache Parquet, it supports efficient compression, predicate pushdown, columnar reads, and parallel processing. GeoParquet integrates seamlessly with modern analytical engines such as DuckDB, Arrow, Spark, and cloud-native query systems, making it ideal for large-scale spatial analysis and reproducible research pipelines.
	\item \textbf{FlatGeobuf} \texttt{(.fgb)}\\
	FlatGeobuf is a high-performance binary format designed for fast spatial access and streaming. It supports built-in spatial indexing and partial reads, enabling efficient use in desktop GIS software (including QGIS) and server-side applications. FlatGeobuf provides functionality comparable to proprietary geodatabases while remaining fully open and single-file, making it well suited for both archival storage and high-performance workflows.
	\item \textbf{Keyhole Markup Language} \texttt{(.kml)}\\
	KML is included primarily for accessibility and outreach. It is widely supported by consumer mapping platforms such as Google Earth and Google Maps, allowing non-technical users, journalists, educators, and civil society organisations to explore the boundaries without specialised GIS software.
\end{enumbul}
We intentionally exclude two Esri-regulated formats, namely Shapefile and File Geodatabase. Shapefiles require a minimum of four companion files to function correctly (\texttt{.shp}, \texttt{.shx}, and \texttt{.dbf} are mandatory, and \texttt{.prj} is required to specify a coordinate reference system), while File Geodatabases are directory-based and governed by proprietary specifications. In contrast, the five formats provided here are open, single-file, and self-describing, making them suitable for long-term archival. We are confident that any contemporary, industry-grade geospatial software can natively support at least one of these formats.

\textbf{Naming convention of delimitation files}

All delimitation files are named as follows:

\begin{centering}

	\texttt{\{unit\_of\_review\}\_\{effective\_year\}\_\{area\_type\}}

\end{centering}

Our naming convention is specifically designed to ensure that users find it intuitive to understand the year in which a boundary set came into legal effect. The case of Sabah's most recent delimitation illustrates this. The EC's website documents the 6th delimitation exercise for Sabah---with the number of Parliaments maintained at 25, and the number of DUNs increased from 60 to 73---as having occurred in 2017. However, this may mislead users into believing that the new boundaries were in effect for the 14th federal general election in 2018, and the concurrently-held 13th Sabah state election. In actuality, because the Prime Minister did not table the EC's recommendations for approval until 2019, the first instance where the new boundaries were in effect was the 14th Sabah state election in 2020. Therefore, even though the EC chooses to enumerate delimitations based on the year in which they concluded their exercise, we argue that the naming convention should capture actual use rather than bureaucratic milestones. Consequently, we denote the most recent delimitation for Sabah as:

\begin{centering}

	\texttt{sabah\_2019\_parlimen}\\
	\texttt{sabah\_2019\_dun}

\end{centering}

That having been said, users should be aware that there are some delimitations---such as the 2018 delimitation of Peninsular Malaysia---which were concluded and approved by Parliament in sufficient time to be used for an election in the same year. Conversely, there are delimitations---such as the 1994 delimitation of Sabah---which were not in effect for the election held in the same year. Consequently, we recommend that users refer strictly to the \texttt{delims\_to\_elections.csv} file when implementing any workflow which assembles election-time maps from the underlying delimitations.

Users should also note that we do not provide a dedicated file covering the special delimitation of 2001 which excised the newly-declared Federal Territory of Putrajaya from the Parliament of Sepang and DUN of Dengkil. This was a deliberate practical choice on our part, because there were no relevant elections of any kind between 2001 and the general election of 2004, which marked the first instance of Putrajaya being contested as a separate electoral area.

\section*{Technical Validation}

We conducted a series of programmatic checks to validate geometric correctness and compliance with the formal rules governing electoral delimitation in Malaysia. These checks were designed to ensure internal consistency, analytical reliability, and interoperability with election results data, rather than to certify boundaries for operational or legal use.

\subsection*{Geometric validation}

All constituency geometries were subjected to standard geometric validation checks commonly applied in geospatial analysis. Specifically, we ensured that:
\begin{enumbul}
	\item all polygons are closed;
	\item all polygons are non-self-intersecting;
	\item all polygons are free of invalid geometries;
	\item all adjacent constituencies align perfectly, with no `holes' or overlaps;
	\item all non-contiguous constituencies are represented as MultiPolygon geometries with identical identifying attributes across component polygons.
\end{enumbul}
These geometric checks are standard practice in the construction of analytical boundary datasets. In the context of our work, they serve an additional role as a safeguard against geo-referencing drift and tracing error during manual digitisation, particularly for early delimitation exercises that relied on printed maps rather than native digital sources.

\subsection*{Compliance with Malaysian delimitation rules}

In addition to geometric correctness, we validated constituency boundaries against the institutional rules governing electoral delimitation in Malaysia. These checks reflect legal and administrative constraints rather than cartographic considerations:

\begin{enumbul}
	\item All Parliaments are exact unions of their component DUNs, with no loss or duplication of area during aggregation;
	\item Each Parliament is linked to exactly one state;
	\item Each DUN is linked to exactly one Parliament and one state.
\end{enumbul}

These constraints were enforced programmatically and ensure that the hierarchical structure of Malaysian electoral geography is preserved across all delimitation exercises.

\subsection*{External validation}

Once all geometric and institutional compliance checks were satisfied, we conducted three additional validation procedures to assess the quality of the dataset against independent authoritative sources.

First, we verified that the number of constituencies in each state and for each delimitation exercise exactly matched the figures explicitly stated in official EC documentation (Figures \ref{fig:nparlimen} and \ref{fig:ndun}). This ensured completeness of coverage, and also provided a guard against inadvertent omission or duplication of constituencies across delimitation cycles.

\begin{figure*}[!htb]
	\centering
	\caption{Computed area vs.\ area listed in delimitation reports}
	\vspace{0.14in}
	\makebox[\textwidth][c]{
		\includegraphics[width=1.25\linewidth]{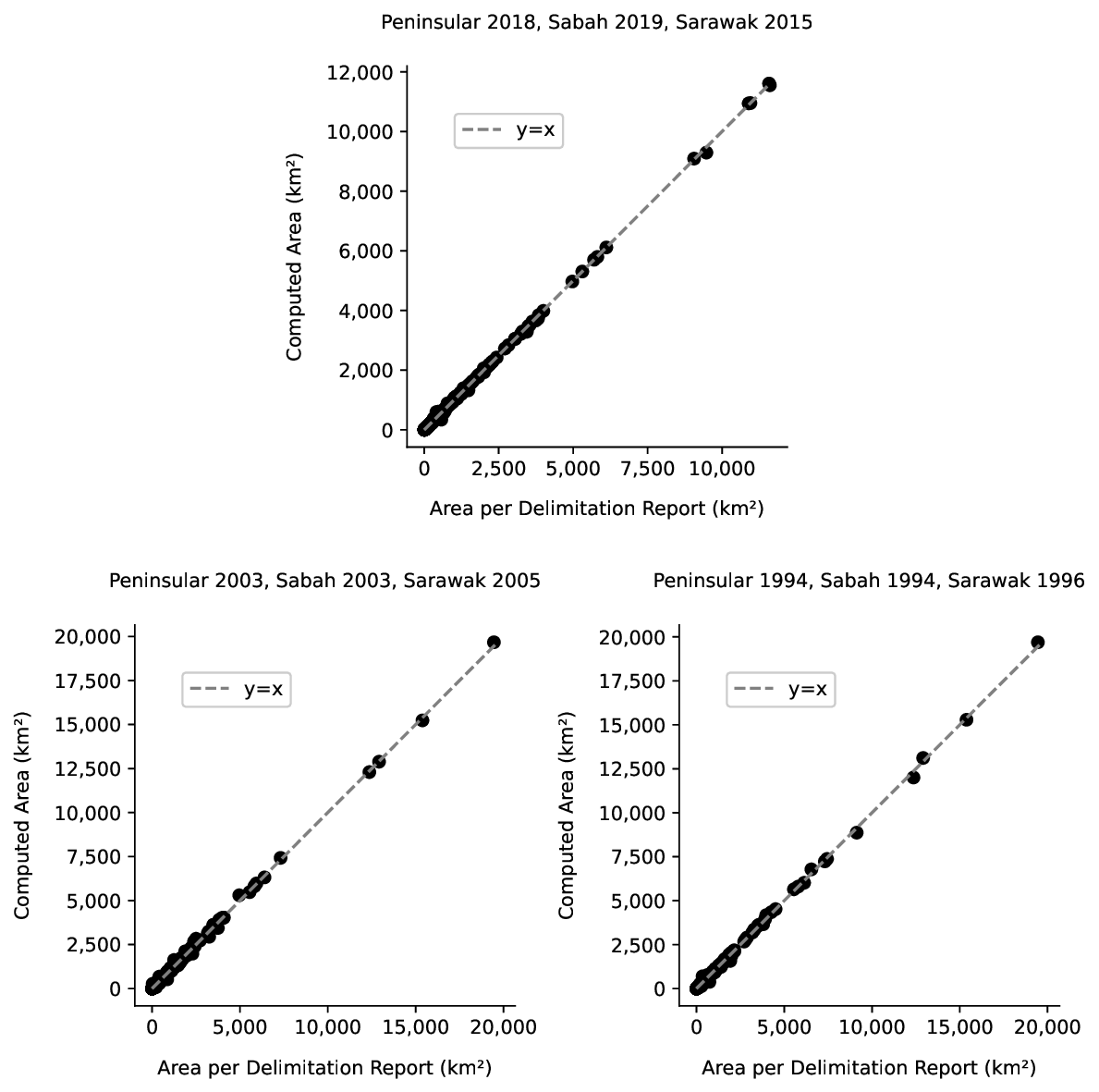}
	}
	\label{fig:errors}
\end{figure*}

Second, for each unit of review, we computed the land area of each DUN constituency and compared it against the official constituency area statistics in delimitation reports for all 9 delimitation exercises from 1994 onwards; area statistics were not published in delimitation reports for the first 10 exercises from 1954 to 1987. We conducted this exercise at the DUN level, because all Parliaments are exact aggregations of their component DUNs, making DUN-level validation both sufficient and higher-resolution. After computing the area of each DUN using our digitised maps, we plotted the computed areas against the officially reported areas (Figure \ref{fig:errors}). Visually, almost all constituencies lie on the $y=x$ line denoting perfect correspondence. Numerically, we carefully analysed the deviations, and concluded that the area of our digitised constituencies falls within $\pm$5\% of the reported figures, except for two classes of constituencies:
\begin{enumbul}
	\item Small constituencies with a land area below 20 km$^2$, such as Komtar in Penang (2 km$^2$), Bandar Hilir in Melaka (5 km$^2$), or Temiang in Negeri Sembilan (6 km$^2$). Official statistics are rounded to the nearest integer, implying that the rounding interval (±0.5 km$^2$) alone can induce an error of 5\% or greater for these constituencies, even before accounting for cartographic generalisation.
	\item Coastal constituencies, such as Kelaboran in Kelantan, Rungkup in Perak, or Pelabuhan Klang in Selangor. The EC's maps are not entirely clear about the exact extent of sea area included in the constituency's area (and may not necessarily be consistent with official jurisdictional maps published by government mapping agencies), thus necessitating some judgment on our part about how to `close' the boundaries of such constituencies such that they have a finite area.
\end{enumbul}
Outside of these clearly defined cases, there are no instances in which the area of a digitised constituency deviates from the published EC figure by more than 5\%. Given our explicitly stated design choice to prioritise analytical correctness over cadastral precision (see Methods), and the lower resolution of historical source material for earlier delimitations, this level of agreement is highly satisfactory. Importantly, this comparison constitutes a strong form of external validation, as it benchmarks the reconstructed geometries against independently published administrative statistics that were not used at any stage of the digitisation process. Because the same tracing and simplification procedures were applied uniformly across all delimitations, this validation provides compelling evidence for the reliability of earlier reconstructed boundaries for which no official area statistics exist.

Third, we verified that every election-time boundary set we generated matched perfectly against the first component of the Malaysian Election Corpus. Since the election results---which were themselves rigorously validated and peer-reviewed---were compiled independently of the boundaries, this check verified that the hierarchical attributes (state, Parliament, DUN) we assigned to each polygon were free of any errors, especially since these were entered manually during the boundary tracing process.

\section*{Usage Notes}

The Malaysian Election Corpus (MECo) is designed to support a wide range of use cases, from rigorous empirical research to exploratory analysis and public-facing data journalism. As the second component of MECo (MECo-2), this work provides the geospatial foundation required to analyse Malaysian elections, complementing the election results data\cite{thevesh2025meco} already released (MECo-1). In the academic realm, we anticipate that this collection will unlock a wave of new research in electoral studies, political geography, and public policy, particularly on issues such as malapportionment, gerrymandering, and spatial patterns of political competition.

We encourage users to use this work in conjunction with MECo-1. All constituency names and identifiers are fully consistent across both components, thus allowing election results and outcomes to be joined directly to boundary geometries without any need for manual reconciliation. Together, the two components support both cross-sectional and longitudinal analysis of Malaysian elections. Beyond election results, the boundaries can also be joined with other spatially referenced datasets---such as census data, administrative statistics, or alternative data---to support broader analyses of the political landscape.

The data can also be interactively explored via \href{https://electiondata.my}{ElectionData.MY}. In addition to improving accessibility for non-technical users, the site serves as a practical demonstration of how the datasets in MECo-1 and MECo-2 can be combined to visualise electoral outcomes, delimitations, and the impact of redelineation. All visualisations presented on ElectionData.MY are generated directly from the datasets released in MECo-1 and MECo-2, and are reproducible using the underlying data.

Finally, while rigorous validation has been applied (see Technical Validation), this collection remains as a living resource. Users are encouraged to report issues or propose improvements via the GitHub repository, which provides full transparency of changes between releases.

\section*{Data Availability}

All datasets described in this paper are published on Harvard Dataverse\cite{meco2_dataset} under a CC0 license. For convenience, the exact same datasets are also mirrored in repositories on Zenodo\cite{meco2_codebase} and \href{https://github.com/Thevesh/paper-meco-maps}{GitHub}, together with the code used for processing, validation, and analysis. We use GitHub to version-control active development and maintenance, with substantial updates released on Zenodo for long-term archival.

\section*{Code Availability}

Delimitations were drawn by hand using QGIS. All processing and validation were done in Python, except for cartogram-generation, which was done in R. The full source code is publicly available under a CC0 license via Zenodo\cite{meco2_codebase} and \href{https://github.com/Thevesh/paper-meco-maps}{GitHub}.

\begingroup
\raggedright
\linespread{1.1}
\small
\let\oldthebibliography\thebibliography
\bibliography{mybib}

@data{meco2_dataset,
  author = {Thevananthan, Thevesh and Prakash Chacko, Danesh},
  publisher = {Harvard Dataverse},
  title = {{The Malaysian Election Corpus (MECo): Electoral Maps and Cartograms since 1954}},
  UNF = {UNF:6:gfjrObr8AOVbJVBxOEFx1g==},
  year = {2025},
  version = {V1},
  doi = {10.7910/DVN/DVFK54},
  howpublished = {\href{https://doi.org/10.7910/DVN/DVFK54}{doi.org/10.7910/DVN/DVFK54}}
 }

@software{meco2_codebase,
  author       = {Thevesh Theva and Danesh Prakash Chacko},
  title        = {Thevesh/paper-meco-maps},
  year         = 2025,
  publisher    = {Zenodo},
  doi          = {10.5281/zenodo.18093017},
  howpublished = {\href{https://doi.org/10.5281/zenodo.18093017}{doi.org/10.5281/zenodo.18093017}},
 }

@article{openshaw1984modifiable,
  title={The modifiable areal unit problem},
  author={Openshaw, Stan},
  journal={Concepts and techniques in modern geography},
  year={1984},
  publisher={GeoBooks}
 }

@article{samuels2001value,
 title={The value of a vote: {Malapportionment} in comparative perspective},
 author={Samuels, David and Snyder, Richard},
 journal={British Journal of Political Science},
 volume={31},
 number={4},
 pages={651--671},
 year={2001},
 publisher={Cambridge University Press}
 }

@data{clea,
  author       = {Kollman, Ken and Hicken, Allen and Caramani, Daniele and Backer, David and Lublin, David},
  title        = {{Constituency-Level Elections Archive (CLEA)}},
  publisher    = {University of Michigan},
  year         = {2024},
  version      = {Ongoing},
  doi          = {10.17616/R33S92},
  howpublished = {\href{https://doi.org/10.17616/R33S92}{doi.org/10.17616/R33S92}},
 }

@article{thevesh2025meco,
  title={{The Malaysian Election Corpus (MECo)}: Federal and State-Level Election Results from 1955 to 2025},
  author={Thevananthan, Thevesh},
  journal={arXiv preprint},
  year={2025},
  doi = {arXiv:2505.06564},
  howpublished = {\href{https://doi.org/10.48550/arXiv.2505.06564}{doi.org/10.48550/arXiv.2505.06564}},
 }

@article{taylor2023canada,
  title={{Canada's Federal Electoral Districts}, 1867--2021: New Digital Boundary Files and a Comparative Investigation of District Compactness},
  author={Taylor, Zack and Lucas, Jack and Kirby, JP and Hewitt, Christopher Macdonald},
  journal={Canadian Journal of Political Science/Revue canadienne de science politique},
  volume={56},
  number={2},
  pages={451--467},
  year={2023}
 }

@book{lewis2013us,
  title={{United States} congressional district shapefiles},
  author={Lewis, Jeffrey B and DeVine, Brandon and Pitcher, Lincoln},
  year={2013},
  publisher={University of California, Los Angeles}
 }

@article{dent1975communication,
  title={Communication aspects of value-by-area cartograms},
  author={Dent, Borden D},
  journal={The American Cartographer},
  volume={2},
  number={2},
  pages={154--168},
  year={1975},
  publisher={Taylor \& Francis}
 }

@article{dorling2011area,
  title={Area cartograms: their use and creation},
  author={Dorling, Daniel},
  journal={The map reader: Theories of mapping practice and cartographic representation},
  pages={252--260},
  year={2011},
  publisher={Wiley Online Library}
 }

@inproceedings{nusrat2016state,
  title={The state of the art in cartograms},
  author={Nusrat, Sabrina and Kobourov, Stephen},
  booktitle={Computer Graphics Forum},
  volume={35},
  number={3},
  pages={619--642},
  year={2016},
  organization={Wiley Online Library}
 }

@article{knowles2002past,
  title={Past time, past place: GIS for history},
  author={Knowles, Anne Kelly},
  journal={(No Title)},
  year={2002}
 }

@misc{long1995atlas,
  title={Atlas of historical county boundaries},
  author={Long, John H},
  year={1995},
  publisher={JSTOR}
 }
\endgroup

\section*{Acknowledgements}
TT thanks Rosmadi Fauzi and Zulkanain Abdul Rahman for their helpful feedback on the manuscript. DPC thanks Tindak Malaysia team member---S. M. Sabri---for their assistance with the digitisation of selected boundary sets. Finally, we thank the EC for granting access to physical election and delimitation reports via \textit{Pusat Sumber SPR}.

\section*{Author Contributions}
TT and DPC both contributed equally to this work.

\section*{Funding}

This work did not receive any specific funding.

\section*{Competing Interests}
The authors declare no competing interests. Neither author is affiliated with any political party.

\end{document}